\documentclass[useAMS]{mn2e}

\usepackage{latexsym,graphicx}

\usepackage{color}

\newcommand\kms{{\rm\,km\,s^{-1}}}

\newcommand\hii{H\,{\sc ii} \,}

\newcommand{\MC}{\multicolumn}

\def\apgt{\ {\raise-.5ex\hbox{$\buildrel>\over\sim$}}\ }
\def\aplt{\ {\raise-.5ex\hbox{$\buildrel<\over\sim$}}\ }

\title[Putative supernova remnant around SXP\,1323]{Discovery of a putative supernova remnant around the long-period X-ray 
pulsar SXP\,1323 in the Small Magellanic Cloud}
\author[V. V.~Gvaramadze, A. Y.~Kniazev and L. M.~Oskinova]
       {V. V.~Gvaramadze$^{1,2}$\thanks{E-mail: vgvaram@mx.iki.rssi.ru}
     A. Y.~Kniazev,$^{3,4,1}$ L. M.~Oskinova$^{5,6}$ \\
    $^{1}$Sternberg Astronomical Institute, Lomonosov Moscow State University, Universitetskij Pr. 13, Moscow 119992, Russia\\
    $^{2}$Space Research Institute, Russian Academy of Sciences, Profsoyuznaya 84/32, Moscow 117997, Russia \\
    $^{3}$South African Astronomical Observatory, PO Box 9, 7935 Observatory, Cape Town, South Africa \\
    $^{4}$Southern African Large Telescope Foundation, PO Box 9, 7935 Observatory, Cape Town, South Africa \\
    $^{5}$Institute for Physics and Astronomy, University Potsdam, 14476 Potsdam, Germany \\
    $^{6}$Kazan Federal University, Kremlevskaya Str 18, Kazan, Russia \\
    }
\begin{document}

\date{Accepted 2019 February 04. Received 2019 February 01; in original form 2019 January 14}


\maketitle

\label{firstpage}

\begin{abstract}
We report the discovery of a circular shell centred on the Be X-ray binary (BeXB) SXP\,1323 in the Small Magellanic 
Cloud (SMC). The shell was detected in an H$\alpha$ image obtained with the Very Large Telescope (VLT). Follow-up 
spectroscopy with the Southern African Large Telescope (SALT) showed that the shell expands with a velocity of 
$\approx100\kms$ and that its emission is due to shock excitation. We suggest that this shell is the remnant of the 
supernova explosion that led to the formation of the SXP 1323's neutron star $\approx40\,000$\,yr ago. SXP\,1323 
represents the second known case of a BeXB associated with a supernova remnant (the first one is SXP\,1062). 
Interestingly, both these BeXBs harbour long period pulsars and are located in a low-metallicity galaxy.
\end{abstract}


\section{Introduction}
\label{sec:int}

\begin{figure*}
\begin{center}
\includegraphics[angle=0,width=0.75\textwidth,clip=]{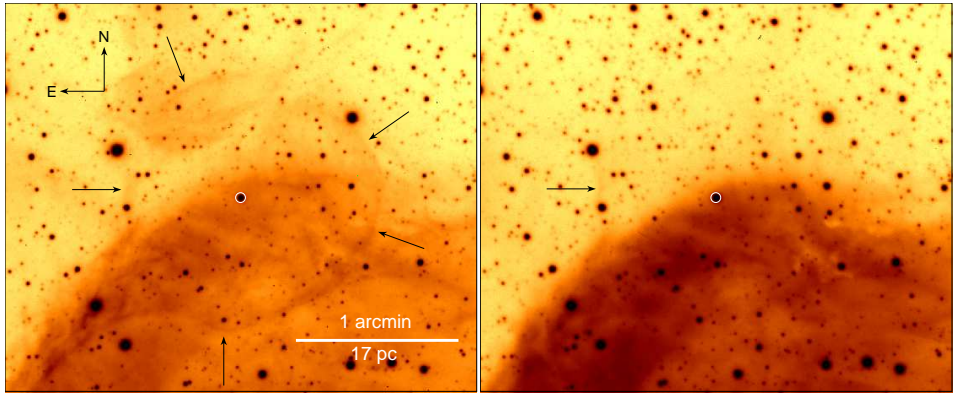}
\end{center}
\caption{VLT H$\alpha$ (left-hand panel) and [O\,{\sc iii}] (right-hand panel) images (obtained by Naz\'e et al. 2003) 
of a region containing the BeXB SXP\,1323 (marked by a circle) and the circular shell around it (indicated by arrows). 
The bright emission to the south of SXP\,1323 is the foreground \hii region N76. The orientation and the scale of 
the images are the same. 
}
\label{fig:shell}
\end{figure*}

Compact objects in X-ray binaries were born in supernova (SN) explosions, however, up to date, only four post-SN binaries 
were found within supernova remnants (SNRs). Three of them are located in the Milky Way. These are the microquasars SS\,433 
(Geldzahler, Pauls \& Salter 1980) and Cir\,X-1 (Heinz et al. 2013), and the pre-low-mass X-ray binary [GV2003]\,N 
(Gvaramadze et al. 2017). The fourth system is the Be X-ray binary (BeXB) SXP\,1062 in the Small Magellanic Cloud (SMC) 
(H\'enault-Brunet et al. 2012; Haberl et al. 2012). There are also two candidate post-SN binaries associated with the 
Galactic SNR RCW\,103 (Bhadkamkar \& Ghosh 2009) and the SNR DEM\,L241 in the Large Magellanic Cloud (Seward et al. 2012), 
but their binary status still have to be confirmed.

In this Letter, we report the discovery of a circular shell around the BeXB SXP\,1323, which we interpret as a SNR produced 
by the same SN explosion that created the SXP\,1323's neutron star. SXP\,1062 and SXP\,1323, therefore, represent the only
two known examples of BeXBs associated with SNRs. Interestingly, both these systems are located in the 
low-metallicity galaxy and the spin periods of their neutron stars are, respectively, the third and second longest known in 
the SMC. In Section\,\ref{sec:she}, we review a relevant information on SXP\,1323 and present Very Large Telescope (VLT) 
H$\alpha$ and [O\,{\sc iii}] images of the shell around this BeXB. Spectroscopic observations of the shell and SXP\,1323 
are described in Section\,\ref{sec:obs}. The obtained results are presented in Section\,\ref{sec:res} and discussed in 
Section\,\ref{sec:dis}.

\section{SXP\,1323 and a circular shell around it}
\label{sec:she}

SXP\,1323 is one of the many high-mass X-ray binaries (HMXBs) in the Small Magellanic Cloud (SMC). Like 
the majority of them,  SXP\,1323 is a BeXB, i.e. consists of an accreting neutron star and a Be donor star. This 
object was identified as a candidate BeXB by Haberl \& Sasaki (2000), who found the emission line star [MA93]\,1393 
(Meyssonnier \& Azzopardi 1993) at the position of the {\it ROSAT} source RXJ0103.6$-$7201. The BeXB status of SXP\,1323 
was confirmed with {\it XMM-Newton} observations, leading to the detection of 1323\,s X-ray pulsations (Haberl \& Pietsch 
2005), which makes SXP\,1323 one of the longest period X-ray pulsars in the SMC known to date (Haberl \& Sturm 2016). 

Follow-up optical spectroscopy of [MA93]\,1393 showed that it is a B0e\,III--V star (McBride et al. 2008), while analysis 
of optical and X-ray light curves of the system revealed its orbital period of $26.2$\,d (Schmidtke \& Cowley 2006; Carpano, 
Haberl \& Sturm 2017). This period is much shorter than what might expect from the empirical pulse/orbital period relationship 
inherent to HMXBs (Corbet 1984, 1986), meaning that the rotational properties of SXP\,1323 are peculiar. Detection of a 
circular shell around SXP\,1323 adds to the singularity of this BeXB.

We found the shell around SXP\,1323 in fig.\,2 of Naz\'e et al. (2003), which shows an H$\alpha$ image of the \hii 
region N76 (of angular diameter of $\approx4$ arcmin) in the north of the SMC. 
On this image one can also distinguish a circular shell of angular diameter of 1.5 arcmin located 
at the northern edge of and partially overlapped with N76 (see Fig.\,\ref{fig:shell}). To our surprise, we 
found no mention of this shell in Naz\'e et al. (2003), while inspection of the SIMBAD data 
base (http://simbad.u-strasbg.fr/simbad/) revealed that the shell is centred on the BeXB SXP\,1323.

The H$\alpha$ image presented by Naz\'e et al. (2003) was obtained with the FOcal Reducer and low dispersion 
Spectrograph instrument (FORS1) on the VLT in 2002. There were also obtained images through several other filters, 
such as [O\,{\sc iii}] (5001 \AA), He\,{\sc i} (5876 \AA) and He\,{\sc ii} (4684 \AA) (see fig.\,4 in Naz\'e et 
al. 2003 for all these images). Inspection of the FORS1 data (http://www.archive.eso.org) showed that the entire shell 
around SXP\,1323 is visible only in the H$\alpha$ line, and that it could also be partially discerned in the 
[O\,{\sc iii}] image, where it appears as an almost straight filament to the east of SXP\,1323 (Fig.\,\ref{fig:shell}).
At the distance to the SMC of 60 kpc (Hilditch, Howarth \& Harries 2005), the linear diameter of the shell is 
$\approx26$ pc.

The central location of SXP\,1323 within the shell suggests that both objects are physically associated
with each other, while the large linear size of the shell (not typical of circumstellar nebulae; e.g.,
Lozinskaya 1992; Chu 2003) points to the possibility that the shell is a SNR. To test this hypothesis, we obtained 
long-slit spectra of the shell, as described in the next section.

\section{SALT spectroscopic observations}
\label{sec:obs}

\begin{table*}
\caption{Journal of the SALT RSS observations of the shell around SXP\,1323.}
\label{tab:log}
\begin{tabular}{llccccccc} \hline
Date & Grating & Exposure & Spectral scale & Spatial scale & PA & Slit & Seeing & Spectral range \\
 &  & (sec) & (\AA\,pixel$^{-1}$) & (arcsec\,pixel$^{-1}$) & ($\degr$) & (arcsec) & (arcsec) & (\AA) \\
 \hline
2011 July 31    & PG900 & 300$\times$4  & 0.97  & 0.255 & 0    &  1.25 & 1.6 & 4200$-$7300 \\
2018 December 20& PG900 & 600$\times$1  & 0.97  & 0.510 & 87   &  1.25 & 1.7 & 3765$-$6900 \\
2018 December 20& PG2300 & 1200$\times$1 & 0.26  & 0.510 & 87   &  2.00 & 1.7 & 6080$-$6870 \\
2018 December 23& PG2300 & 1200$\times$1 & 0.26  & 0.510 & 87   &  2.00 & 1.8 & 6080$-$6870 \\
  \hline
\end{tabular}
\end{table*}

The spectra of the shell around SXP\,1323 were obtained in 2011 and 2018 with the Robert Stobie Spectrograph (RSS; 
Burgh et al. 2003; Kobulnicky et al. 2003) mounted on the Southern African Large Telescope (SALT; Buckley, Swart 
\& Meiring 2006; O'Donoghue et al. 2006). The observations were carried out in the long-slit spectroscopy mode
with the slit placed on SXP\,1323.

The first two spectra were obtained using the PG900 grating with the spectral resolution FWHM of 4.7 \AA.
In 2011 the spectroscopic slit was oriented in the south-north direction, while in 2018 it was oriented at the 
position angle PA=87\degr \, to cross the [O\,{\sc iii}] filament on the eastern side of the shell. We will refer to 
these spectra as the `low-resolution spectra'. Besides the PG900 grating, the PG2300 grating was also used to study 
the velocity distribution along the shell. The spectra obtained with this grating have the spectral resolution 
FWHM of 1.8~\AA. We will refer to them as the `high-resolution spectra'. Spectrophotometric standards were observed 
at the same spectral setups during nearest twilights as a part of SALT calibration plan. For the log of observations 
see Table\,\ref{tab:log}.

Primary reduction of the RSS data was done with the SALT science pipeline (Crawford et al. 2010). The subsequent 
long-slit data reduction was carried out in the way described in Kniazev et al. (2008). Absolute flux calibration
is not feasible with SALT because the unfilled entrance pupil of the telescope moves during the observation. However, 
a relative flux correction to recover the spectral shape was done using the observed spectrophotometric standard.
The resulting reduced RSS spectra of the shell and its central star are presented and discussed in the next section.

\section{Results of spectroscopy}
\label{sec:res}

\begin{figure}
\begin{center}
\includegraphics[angle=0,width=0.4\textwidth,clip=]{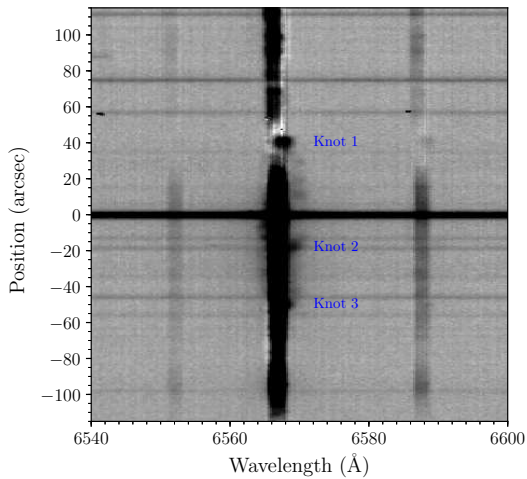}
\end{center}
\caption{A part of the 2D high-resolution spectrum of the shell (PA=87\degr), showing the H $\alpha$ and [N\,{\sc ii}] 
$\lambda\lambda$6548, 6584 emission lines. East is up and west is down. See the text for details.}
\label{fig:2D}
\end{figure}
%
\begin{figure}
\begin{center}
\includegraphics[angle=0,width=0.85\columnwidth,clip=]{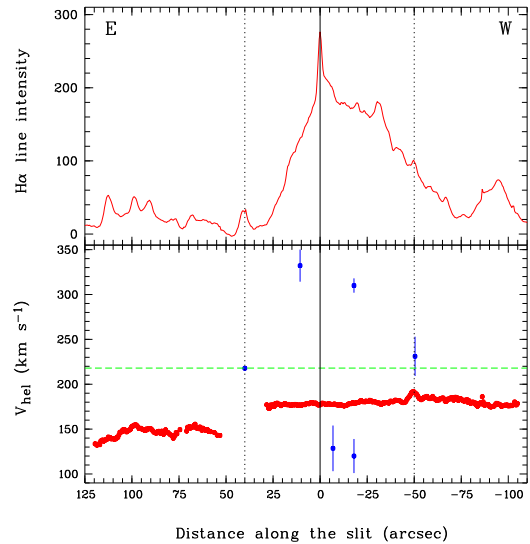}
\end{center}
\caption{Upper panel: H$\alpha$ line intensity along the RSS slit (PA=87\degr). The E--W direction of the slit 
is shown. Bottom panel: H$\alpha$ heliocentric radial velocity distribution along the slit. The (blue) dots with
error bars show the heliocentric radial velocity of the shell measured at several positions in the 2D spectrum 
(Fig.\,\ref{fig:2D}). The dashed horizontal line corresponds to the systemic velocity of the shell ($\approx220 \, 
\kms$). The solid vertical line shows the position of SXP\,1323, while the dotted vertical lines mark the edges of 
the shell. The thick noisy (red) line shows the heliocentric radial velocity distribution for the \hii region N76 
($V_{\rm hel}\approx170 \, \kms$) and the background H$\alpha$ emission to the east of the shell ($V_{\rm hel}\approx150
\, \kms$).} 
\label{fig:Ha}
\end{figure}

Fig.\,\ref{fig:2D} presents a portion of the high-resolution spectrum of the shell in the region of the H$\alpha$ 
line, showing the presence of expanding shell, whose eastern and western rims (located, respectively, at +40 arcsec 
and $-50$ arcsec) are labelled as knot\,1 and knot\,3. There are also several brightness enhancements within the shell 
of which the brightest one is labelled as knot\,2. Using the 2D spectrum, we plot the distributions of the H$\alpha$ 
emission line intensity and heliocentric radial velocity, $v_{\rm hel}$, along the slit (see Fig.\,\ref{fig:Ha}). From 
the bottom panel of Fig.\,\ref{fig:Ha} if follows that the shell expands with a velocity of $96\pm17\kms$ and that its 
systemic velocity is $219\pm17\kms$. The latter velocity is offset with respect to the systemic velocity of the \hii 
region N76 of $\approx180\kms$ (see Fig.\,\ref{fig:Ha}) and the mean radial velocity of massive stars in the 
SMC of $\approx170\kms$ (Evans \& Howarth 2008), suggesting that the SN progenitor was a runaway star. Fig.\,\ref{fig:Ha} 
also shows that SXP\,1323 is offset from the geometric centre of the shell in the eastern direction by about 5 arcsec.

\begin{figure}
\begin{center}
\includegraphics[angle=-90,width=1\columnwidth,clip=]{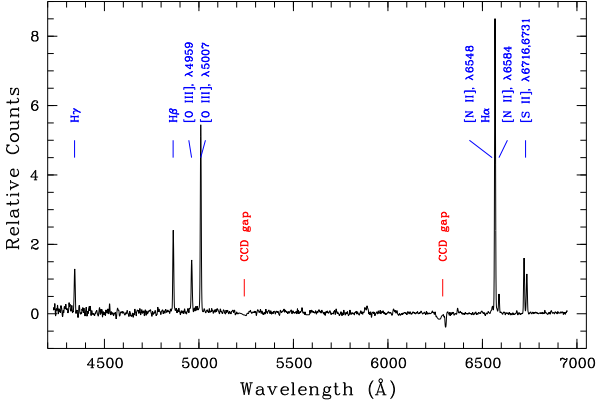}
\end{center}
\caption{1D low-resolution spectrum of the northern rim of the shell (PA=0\degr). The major emission lines are marked.} 
\label{fig:1D}
\end{figure}

Fig.\,\ref{fig:1D} shows the 1D low-resolution spectrum of the northern rim of the shell (PA=0\degr). The major emission 
lines detected in this and the second low-resolution spectrum (PA=87\degr) were measured using the programs described in 
Kniazev et al. (2004), and their reddening-corrected line intensity ratios, $I(\lambda)/I($H$\beta)$, are given in 
Table\,\ref{tab:int} along with the logarithmic extinction coefficient $C$(H$\beta$), the colour excess $E(B-V)$, and 
the electron number density $n_{\rm e}$([S\,{\sc ii}]) measured from the [S\,{\sc ii}] $\lambda\lambda$6716, 6731 line 
ratio. For the sake of completeness, to this table we also added the line intensities in the high-resolution 
spectrum of knot\,2. The derived values of $E(B-V)$ are larger than the colour excess of N76 of $0.12\pm0.02$ mag 
(derived from the 2011's spectrum), meaning that the shell is located behind this \hii region.

\begin{table}
\centering{
\caption{Columns 2 and 3: relative intensities of major emission lines in the low-resolution spectra of the
northern (PA=0\degr) and eastern (PA=87\degr) rims of the shell. Column 4: intensities of lines in the
high-resolution spectrum of knot\,2.}
\label{tab:int}
\begin{tabular}{lccc} \hline
\rule{0pt}{10pt}
$\lambda_{0}$(\AA) Ion      & $I(\lambda)/I({\rm H}\beta)$ & $I(\lambda)/I({\rm
H}\beta)$ & $I(\lambda)$ \\
\hline
 & \MC{1}{c}{PA=0\degr} & \MC{1}{c}{PA=87\degr}    &  \MC{1}{c}{PA=87\degr}    \\
4340\ H$\gamma$\            &  0.54$\pm$0.03 & ---            & --- \\
4861\ H$\beta$\             &  1.00$\pm$0.01 & 1.00$\pm$0.07  & --- \\
4959\ [O\ {\sc iii}]\       &  0.66$\pm$0.02 & 1.26$\pm$0.06  & --- \\
5007\ [O\ {\sc iii}]\       &  2.19$\pm$0.06 & 3.49$\pm$0.14  & --- \\
6563\ H$\alpha$\            &  2.88$\pm$0.08 & 2.88$\pm$0.15  & 2.08$\pm$0.26 \\
6584\ [N\ {\sc ii}]\        &  0.18$\pm$0.01 & 0.19$\pm$0.02  & 0.16$\pm$0.04 \\
6717\ [S\ {\sc ii}]\        &  0.53$\pm$0.02 & 0.52$\pm$0.03  & 0.51$\pm$0.06 \\
6731\ [S\ {\sc ii}]\        &  0.38$\pm$0.01 & 0.33$\pm$0.02  & 0.35$\pm$0.06 \\
$C$(H$\beta$)\,(dex)        & 0.29$\pm$0.03  & 0.25$\pm$0.05  & --- \\
$E(B-V)$\,(mag)             & 0.20$\pm$0.02  & 0.17$\pm$0.03 & --- \\
$n_{\rm e}$([S\,{\sc ii}])\,(cm$^{-3}$) & 26$^{+55} _{-16}$ & --- & --- \\
\hline
\end{tabular}
}
\end{table}

\begin{figure}
\begin{center}
\includegraphics[angle=-90,width=0.8\columnwidth,clip=]{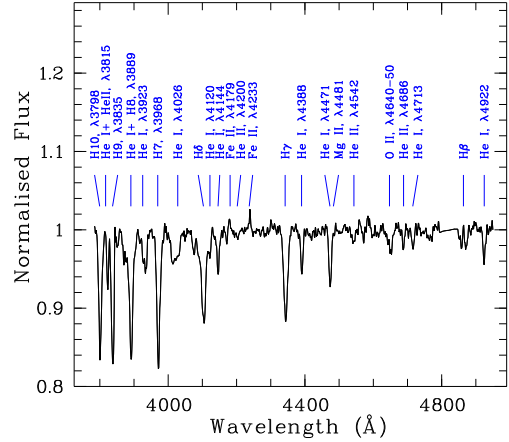}
\includegraphics[angle=-90,width=0.8\columnwidth,clip=]{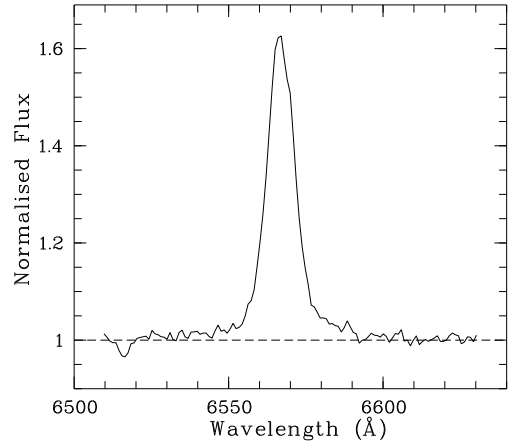}
\end{center}
\caption{Upper panel: a blue part of the low-resolution RSS spectrum of [MA93]\,1393 obtained on 2018 December 20.
Bottom panel: H$\alpha$ line in the same spectrum.} 
\label{fig:star}
\end{figure}

Fig.\,\ref{fig:star} shows the normalized spectrum of [MA93]\,1393 --- the donor star in SXP\,1323, which is dominated 
by the H$\alpha$ emission line. The H$\beta$ absorption line is filled with emission, while other Balmer lines are in
absorption. Two other emission lines detected in the spectrum are those of iron, Fe\,{\sc ii} $\lambda\lambda$4179, 4233. 
The presence of these lines is typical of Be stars. Using the classification criteria from Evans et al. (2004), we derived 
the spectral class of [MA93]\,1393 of B0, while from the photometry of this star (taken from McBride et al. 2008) and the 
absolute magnitude calibration from Walborn (1972), we found that the star is of luminosity class III, which is in a 
good agreement with the results by McBride et al. (2008). The obtained classification is further supported by estimates of 
the effective temperature and gravity of [MA93]\,1393 of $T_{eff}=30\,600\pm1000$\,K and $\log g=3.47\pm0.06$ derived from 
the low-resolution spectra using the {\sc ulyss} program (Koleva et al. 2009) and a medium spectral-resolution library 
(Prugniel, Vauglin \& Koleva 2011), with which we also measured the heliocentric radial velocity of the star of 
$145\pm20\kms$.

We also measured the equivalent width (EW) of the H$\alpha$ line in the spectra of [MA93]\,1393 to check whether it 
follows the orbital period--EW(H$\alpha$) correlation revealed for BeXBs (Reig, Fabregat \& Coe 1997). The obtained 
values of $-4.8\pm0.3$\,\AA \, (2011) and $-8.2\pm0.2$\,\AA \, (2018) nicely fit this correlation and imply the orbital 
period of $\approx20-30$\,d (see fig.\,9 in Haberl \& Sturm 2016) in a good agreement with the observed value of 26.2\,d 
(Carpano et al. 2017). This indicates that the size of the decretion disk around [MA93]\,1393 is set by the feedback from 
the companion neutron star (Reig et al. 1997; Coe \& Kirk 2015).

\section{Discussion}
\label{sec:dis}

Spectroscopic observations of the shell around SXP\,1323 showed that it expands with a velocity of 
$V_{\rm sh}\approx100\kms$, which is typical of mature SNRs (e.g. Lozinskaya 1992). The ratio of the 
combined [S\,{\sc ii}] $\lambda\lambda$6716, 6731 lines against H$\alpha$ of 0.3--0.4 (see Table\,\ref{tab:int}) 
agrees with the SNR interpretation of the shell, although it is at the lower end of a range of values measured for 
SNRs (e.g. Fesen, Blair \& Kirshner 1985). The moderate value of this ratio could be due to photoionization
of the shell and the ambient interstellar medium (ISM) by the donor star in SXP\,1323 and/or by the WN4+O6\,I(f) star 
SMC\,AB\,7 located in the centre of N76 at $\approx30$ pc (in projection) from SXP\,1323 (cf. H\'enault-Brunet et al.
2012).

If the SNR is in the adiabatic phase, then its radius, $R_{\rm SNR}$, can be described by the Sedov (Sedov 
1959) solution:
\begin{equation}
R_{\rm SNR}=\left({2.026E_{\rm SN}\over \rho_{\rm ISM}}\right)^{1/5} t^{2/5} \, ,
\label{eqn:sed}
\end{equation}
where $E_{\rm SN}$ is the kinetic energy produced by the SN explosion, $\rho_{\rm ISM}=1.4m_{\rm H}n_{\rm ISM}$, 
$m_{\rm H}$ is the mass of the hydrogen atom, and $n_{\rm ISM}$ is the number density of the local ISM. In this 
case, the age of the SNR is given by $t_{\rm SNR}=0.4(R_{\rm SNR}/V_{\rm SNR})$, where 
$V_{\rm SNR}=(4/3)V_{\rm sh}$ is the expansion velocity of the SN blast wave. For $R_{\rm SNR}=12.9$ pc and 
$V_{\rm sh}=100\kms$, one finds $t_{\rm SNR}\approx4\times10^4 \, {\rm yr}$. Inserting 
this age in equation (\ref{eqn:sed}), one finds that $E_{51}/n_{\rm ISM}=0.08$, where $E_{51}=E_{\rm SN}/(10^{51} 
\, {\rm erg}$), implying $n_{\rm ISM}\approx13 \, {\rm cm}^{-3}$ for the standard energy of the SN explosion of 
$E_{51}=1$, or $n_{\rm ISM}\approx1.3 \, {\rm cm}^{-3}$ if $E_{51}=0.1$. Both these number densities, however, imply 
that the SNR should already be in the snow-plough phase of its evolution (e.g. Cioffi, McKee \& Bertschinger 1988),
which is in variance with our assumption that it is adiabatic.

To get an idea on $n_{\rm ISM}$, one can use the relationship $n_{\rm ISM}=0.02 \, 
{\rm cm}^{-3} (V_{\rm SNR}/100 \, \kms)^{-2} n_{\rm e}$([S\,{\sc ii}]) (Dopita 1979). Using this equation and 
taking into account that in the snow-plough phase $V_{\rm SNR}\approx V_{\rm sh}$, one finds that $n_{\rm ISM}=0.5-1.6 
\, {\rm cm}^{-3}$ if one adopts $n_{\rm e}$([S\,{\sc ii}])=$25-80 \, {\rm cm}^{-3}$ (see Table\,\ref{tab:int}). Then, 
using equations (3.32) and (3.33) in Cioffi et al. (1988) and the observed radius and expansion velocity of the shell, 
one finds the age of the SNR of $\approx4\times10^4$\,yr (if $E_{51}=1$) or $\approx2.5\times10^4$\,yr (if $E_{51}=0.1$).

These ages appear to be too young for the neutron star in SXP\,1323 to spin-down to its current spin period if it was
born as a millisecond rotator (e.g. Urpin, Konenkov \& Geppert 1998). In this sense, SXP\,1323 is similar to
SXP\,1062, which, like SXP\,1323, is one of the longest spin period X-ray pulsars in the SMC and which up to 
now was the only known BeXB within a SNR (H\'enault-Brunet et al. 2012; Haberl et al. 2012). (The detailed comparison 
of these two systems will be presented elsewhere.) 

It is believed that the paucity of X-ray pulsar/SNR associations is because the time required for the post-SN binary 
to evolve into an X-ray pulsar is much longer than the lifetime of SNRs of $\sim10^5$\,yr. The detection of two SNRs 
around long-period BeXBs challenges the traditional view on the formation time-scales for HMXBs (see, e.g., 
Gonz\'alez-Gal\'an et al. 2018 and references therein) and hints at the possibility that some neutron stars could be 
slow rotators from birth (e.g. Knigge, Coe \& Posiadlowski 2011), which in turn could naturally explain the observed
bimodality in the pulse period distribution of BeXBs (Knigge et al. 2011; Coe \& Kirk 2015). Search for SNRs around 
other BeXBs is needed to understand whether the detection of SNRs around two longest period BeXBs is a simple 
coincidence or has more far-reaching consequences. 

To conclude, we note that the disparity between the systemic velocity of the shell and the heliocentric radial 
velocity of [MA93]\,1393 suggests that the post-SN binary was kicked towards us with a velocity of $\approx74\pm26\kms$. 
The space velocity of SXP\,1323 could be a bit higher if the 5 arcsec (or $\approx1.43$ pc) offset of this system from 
the geometric centre of the SNR is due to a kick in the transverse direction.

\section{Acknowledgements}
This work is based on observations obtained with the Southern African Large Telescope (SALT), programmes 
2010-1-RSA-OTH-001 and 2018-1-MLT-008, and supported by the Russian Foundation for Basic Research grant 19-02-00779. 
AYK acknowledges support from the National Research Foundation (NRF) of South Africa. LMO acknowledges partial 
support by the Russian Government Program of Competitive Growth of Kazan Federal University. This research has made 
use of the SIMBAD data base, operated at CDS, Strasbourg, France.


\begin{thebibliography}{}
%
\bibitem{} Bhadkamkar H., Ghosh P., 2009, A\&A, 506, 1297
\bibitem{} Buckley D. A. H., Swart G. P., Meiring J. G., 2006, in Stepp L. M., ed., Proc. SPIE Conf. Ser. Vol. 6267, 
Ground-based and Airborne Telescopes. SPIE, Bellingham, p. 62670Z
\bibitem{} Burgh E. B., Nordsieck K. H., Kobulnicky H. A., Williams T. B., O'Donoghue D., Smith M. P., Percival J. W., 2003,
in Iye M., Moorwood A. F. M., eds, Proc. SPIE Conf. Ser. Vol. 4841, Instrument Design and Performance for Optical/Infrared
Ground-based Telescopes. SPIE, Bellingham, p. 1463
\bibitem{} Carpano S., Haberl F., Sturm R., 2017, A\&A, 602, A81
\bibitem{} Chu Y.-H., 2003, in van der Hucht K., Herrero A., Esteban C., eds, Proc. IAU Symp. 212, A Massive Star 
Odyssey: From Main Sequence to Supernova. Astron. Soc. Pac., San Francisco, p. 585
\bibitem{} Cioffi D. F., McKee C. F., Bertschinger E., 1988, ApJ, 344, 252
\bibitem{} Coe M. J., Kirk J., 2015, MNRAS, 452, 969
\bibitem{} Corbet R. H. D., 1984, A\&A, 141, 91
\bibitem{} Crawford S. M. et al., 2010, in Silva D. R., Peck A. B., Soifer B. T., Proc. SPIE Conf. Ser. Vol. 7737, Observatory
Operations: Strategies, Processes, and Systems III. SPIE, Bellingham, p. 773725
\bibitem{} Dopita M. A., 1979, ApJS, 40, 455
\bibitem{} Evans C. J., Howarth I. D., 2008, MNRAS, 386, 826
\bibitem{} Evans C. J., Howarth I. D., Irwin M. J., Burnley A. W., Harries T. J., 2004, MNRAS, 353, 601
\bibitem{} Fesen R. A., Blair W. P., Kirshner R. P., 1985, ApJ, 292, 29
\bibitem{} Geldzahler B. J., Pauls T., Salter C. J., 1980, A\&A, 84, 237
\bibitem{} Gonz\'alez-Gal\'an A. et al., 2018, MNRAS, 475, 2809
\bibitem{} Gvaramadze V. V. et al., 2017, Nature Astron., 1, 0116
\bibitem{} Haberl F., Sasaki M., 2000, A\&A, 359, 573
\bibitem{} Haberl F., Sturm R., 2016, A\&A 586, A81 
\bibitem{} Haberl F., Pietsch W., 2005, A\&A, 438, 211
\bibitem{} Haberl F., Sturm R., Filipov\'ic M. D., Pietsch W., Crawford E. J., 2012, A\&A, 537, L1
\bibitem{} Heinz S. et al., 2013, ApJ, 779, 171
\bibitem{} H\'enault-Brunet V. et al., 2012, MNRAS, 420, L13
\bibitem{} Hilditch R. W., Howarth I. D., Harries T. J., 2005, MNRAS, 357, 304
\bibitem{} Kniazev A. Y., Pustilnik S. A., Grebel E. K., Lee H., Pramskij A. G., 2004, ApJS, 153, 429
\bibitem{} Kniazev A. Y. et al., 2008, MNRAS, 388, 1667
\bibitem{} Knigge C., Coe M. J., Posiadlowski P., 2011, Nature, 479, 372
\bibitem{} Kobulnicky H. A., Nordsieck K. H., Burgh E. B., Smith M. P., Percival J. W., Williams T. B., O'Donoghue D., 2003, in Iye M.,
Moorwood A. F. M., eds, Proc. SPIE Conf. Ser. Vol. 4841, Instrument Design and Performance for Optical/Infrared Ground-based Telescopes.
SPIE, Bellingham, p. 1634
\bibitem{} Koleva M., Prugniel P., Bouchard A., Wu Y., 2009, A\&A, 501, 1269
\bibitem{} Lozinskaya T. A., 1992, Supernovae and Stellar Wind in the Interstellar Medium. Am. Inst. Phys., New York
\bibitem{} McBride V. A., Coe M. J., Negueruela I., Schurch M. P. E., McGowan K. E., 2008, MNRAS, 388, 1198
\bibitem{} Meyssonnier N., Azzopardi M., 1993, A\&AS, 102, 451
\bibitem{} Naz\'e Y., Rauw G., Manfroid J., Chu Y.-H., Vreux J.-M., 2003, A\&A, 408, 171
\bibitem{} O'Donoghue D. et al., 2006, MNRAS, 372, 151
\bibitem{} Prugniel P., Vauglin I., Koleva M., 2011, A\&A, 531, 165
\bibitem{} Reig P., Fabregat J., Coe M. J., 1997, A\&A, 322, 193
\bibitem{} Schmidtke P. C., Cowley A. P., 2006, AJ, 132, 919
\bibitem{} Sedov L. I., 1959, Similarity and Dimensional Methods in Mechanics, Academic Press, New York
\bibitem{} Seward F. D. et al., 2012, ApJ, 759, 123
\bibitem{} Urpin V., Konenkov D., Geppert U., 1998, MNRAS, 299, 73
\bibitem{} Walborn N.R., 1972, AJ, 77, 312
%
\end{thebibliography}
\end{document}